\documentclass[aip,amsmath,twocolumn,nobalancelastpage,reprint,
]{revtex4-1}
\pdfoutput=1

\usepackage[caption=false]{subfig}
\usepackage{graphicx}
\usepackage{dcolumn}
\usepackage{bm}
\usepackage{epstopdf}
\usepackage{adjustbox}
\usepackage{pifont}
\usepackage{overpic}
\makeatletter
\@namedef{ver@hobsub-hyperref.sty}{}
\@namedef{ver@refcount.sty}{}
\makeatother
\usepackage[colorlinks=true,citecolor=black,urlcolor=black,linkcolor=black]{hyperref}

\newcommand{\cmark}{\ding{51}}

\begin{document}

\preprint{AIP/2016}

\title{A Reconfigurable Cryogenic Platform for the Classical Control of Scalable Quantum Computers}
\thanks{The extended article has been published in Review of Scientific Instruments. It is available online at \href{https://dx.doi.org/10.1063/1.4979611}{https://dx.doi.org/10.1063/1.4979611}.}

\author{Harald Homulle}
\author{Stefan Visser}
\author{Bishnu Patra}
\affiliation{QuTech, Delft University of Technology, 2628CD Delft, The Netherlands}
\author{Giorgio Ferrari}
\affiliation{Politecnico di Milano, 20133 Milan, Italy}
\author{Enrico Prati}
\affiliation{Istituto di Fotonica e Nanotecnologie, 20133 Milan, Italy}
\author{Fabio Sebastiano}
\author{Edoardo Charbon}
\email{e.charbon@tudelft.nl}
\affiliation{QuTech, Delft University of Technology, 2628CD Delft, The Netherlands}
 
\date{2 February 2016}
\revised{\today}

\begin{abstract}
Recent advances in solid-state qubit technology are paving the way to fault-tolerant quantum computing systems. However, qubit technology is limited by qubit coherence time and by the complexity of coupling the quantum system with a classical electronic infrastructure.

We propose an infrastructure, enabling to read and control qubits, that is implemented on a field-programmable gate array (FPGA). The FPGA platform supports functionality required by several qubit technologies and can operate physically close to the qubits over a temperature range from 4K to 300K. Extensive characterization of the platform over this temperature range revealed all major components (such as LUTs, MMCM, PLL, BRAM, IDELAY2) operate correctly and the logic speed is very stable. The stability is finally concretized by operating an integrated ADC with relatively stable performance over temperature.
\end{abstract}


\maketitle

\section{Introduction}
\noindent In a fault-tolerant quantum computing system based on either electron / hole spin in semiconductors or superconductors, each qubit generally operates at sub-1K temperatures and requires a classical loop capable of reading the state of the qubit and controlling it based on a (localized) decision. To be effective, such error-correcting loops need to perform a complete correction cycle faster than the qubit decoherence time, which, depending on the qubit technology, can vary from ms to $\mu$s \cite{Braakman2013, Kawakami2014, Kim2014, Colless2013, Colless2012, Kalra2014, Muhonen2014, Muhonen2015, Dicarlo2009, Riste2015, Barends2014, Jeffrey2014}. Reading a qubit requires a certain signal-to-noise ratio (SNR) and it should not interfere with the qubit state itself, whereas, to support large systems, error-correcting loops should be physically small and compatible with massively parallel operation.

Future error-correcting loops will likely be implemented in micro-architectures supported by a hardware infrastructure that operates at cryogenic temperatures, so as to ensure proximity to qubits and compactness. Moreover, to avoid time- and energy-consuming cooling cycles, micro-architectures supporting error correction should be programmable and/or reconfigurable. Several technologies exist that could support logic circuits at deep cryogenic temperatures: GaAs high electron mobility transistors (HEMTs) \cite{Chow2014}, rapid single flux quantum (RSFQ) devices \cite{Castellano2006, Mukhanov2011}, and custom semiconductors \cite{Ward2013, Al-Taie2013, Puddy2015, Prager2011}, to name a few. These technologies are currently expensive and generally not scalable, while CMOS processes can leverage a mature infrastructure that is likely to continue improving for several more decades \cite{Prati2014}. To meet the reconfigurability requirement, the obvious choice is an FPGA fabricated in a deep-submicron (DSM) CMOS process, due to its density, versatility, and full compatibility with standard CMOS circuits.

DSM CMOS circuits have been known to operate at deep cryogenic temperatures, down to several hundreds of mK \cite{Ward2013, Al-Taie2013, Puddy2015, Prager2011, Prati2012, Turchetti2016}. Cryogenic FPGAs can operate at 4K as a quantum controller \cite{Hornibrook2015, ConwayLamb2016}. The Spartan 3, Spartan 6, and Artix 7 FPGAs by Xilinx were shown to operate at 4K, however the limits of operability at 4K is still largely unknown.

We explore the regime of operability of a commercial FPGA (Artix 7, Xilinx) to enable a full error-correcting loop for fault-tolerant operation of solid-state qubits. The resulting infrastructure is scalable and versatile, efficiently integrated in an FPGA and operating at low power. In order to implement such a platform, a subset of the components required in error-correction loops were designed and tested. Moreover extensive characterisation of the FPGA performance in cryogenic conditions was carried out. While the target of this study is a specific spin qubit \cite{Kawakami2014, Braakman2013}, the platform can in principle also be used for superconducting qubits \cite{Dicarlo2009, Riste2015} and for other qubits.

\section{An FPGA based control platform}
\begin{figure}[t]
	\centering
		\includegraphics[width=0.45\textwidth]{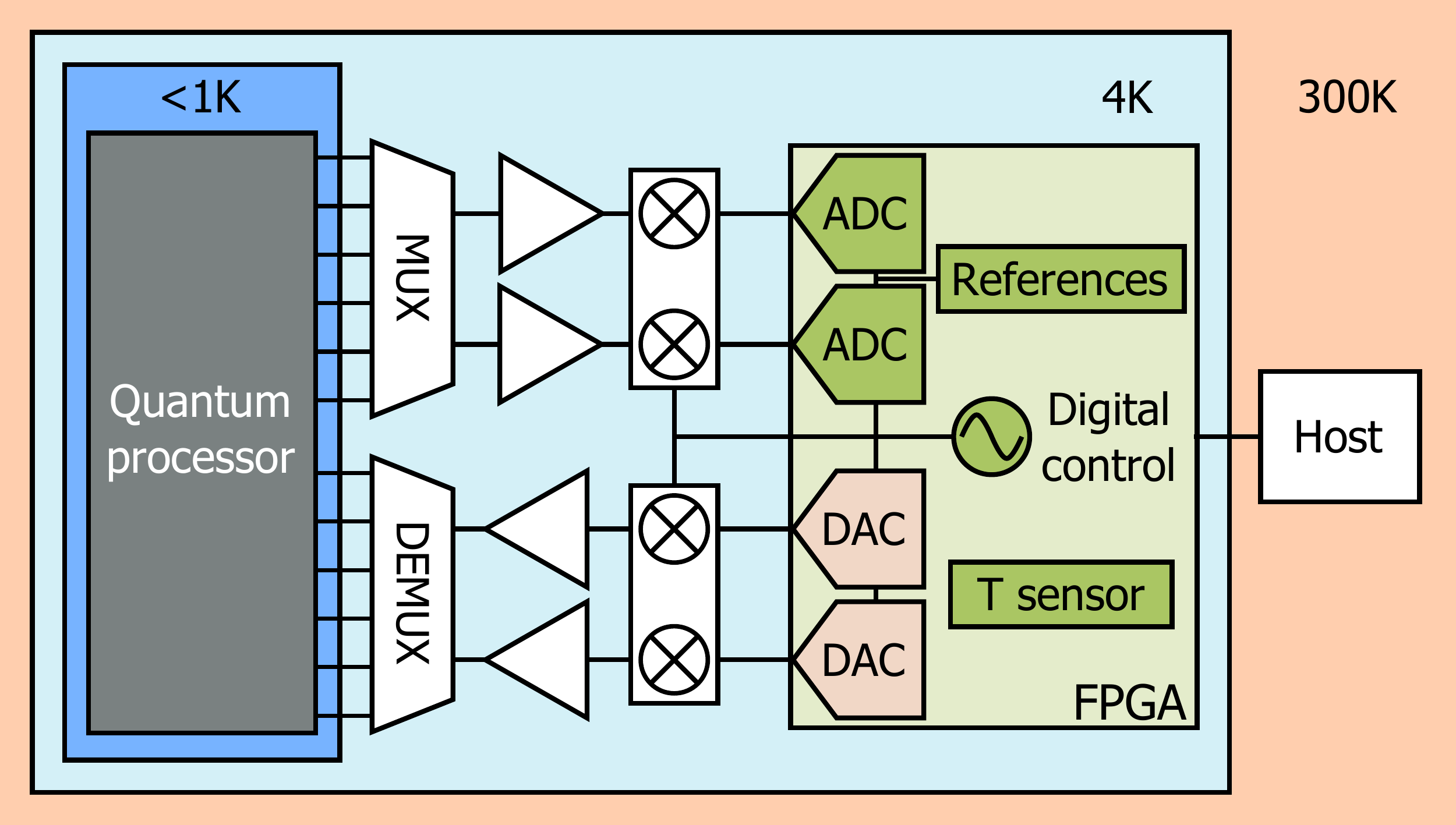}
		\caption{
		Cryogenic control platform implementing the infrastructure supporting error-correcting micro-architectures. An all-digital implementation of the main components and integration inside the FPGA is shown. Although in theory both red and green components can be integrated, in the following the interest is focussed on the green parts. }
		\label{fig:control_platform}
	\end{figure}

\noindent The architecture of the control electronics of quantum devices is shown in \autoref{fig:control_platform}; it comprises multiplexers and demultiplexers, in close proximity to or integrated with the qubits, amplifiers, analog-to-digital converters (ADCs), digital-to-analog converters (DACs), oscillator(s), down- and up-converting mixers, and general-purpose digital logic.  

Operating semiconductor devices at deep cryogenic temperatures is a challenge, due to freeze-out, non-idealities in transistor I-V characteristics, and increased mismatch \cite{Das2014}. While freeze-out effects at 4K are less problematic in DSM CMOS technologies due to the high levels of doping, MOS transistors I-V characteristics exhibit a so-called `kink', which causes elevated current levels at high V$_\text{DS}$ (due to a reduction in the transistors threshold voltage V$_\text{t}$). Furthermore, hysteresis in the drain current when sweeping the drain voltage upwards or downwards is increased at cryogenic temperatures.

To overcome these limitations, we opted for a fully digital approach by way of FPGAs. The redesign of analog circuits adjusted to cryogenic conditions can (at least partially) be avoided, since the large majority of the required components can be implemented as a digital block, as shown in \autoref{fig:control_platform}. 

Analog signals are converted to digital codes at the interface with the FPGA. This data is processed digitally to measure the qubit state and to synthesize the appropriate correction. To avoid further unnecessary signal conversions, and thus maximizing SNR and signal integrity, control signals are generated directly by the FPGA, whenever possible, or by DACs. Although DACs can be integrated in the FPGA, they will not be considered in this work.

Operating an FPGA almost 250K below the standard operating range, is not trivial and it comes with several challenges. 
	
For what concerns power dissipation, when reducing the temperature, power dissipation budgets become increasingly stricter, down to a few watts at 4K, due to limitations in refrigeration technologies. It imposes care in designing custom integrated circuits, and even more so in FPGAs. To fully characterize FPGA performance, the platform is first tested in liquid helium directly, thus maximizing cooling power if compared to closed-loop refrigeration. 

Next, we consider the physical implementation of the platform, needing \textit{ad-hoc} printed circuit boards (PCBs) hosting robust passive components and connectors that can reliably survive several cooling cycles without performance degradation. In our design, we used a minimal number of discrete components certified for operation over a temperature range, wider than industrial standards.

The most critical challenge is the operating condition and the corresponding behaviour of the FPGA. Especially for high performance circuits, such as time-to-digital converters (TDCs) and ADCs, the FPGA behaviour has to be well understood.

\section{A cryogenic FPGA}\label{sec:cryo_results_basic}
\begin{figure}[t]
		\centering
		\begin{overpic}[width=0.49\textwidth]{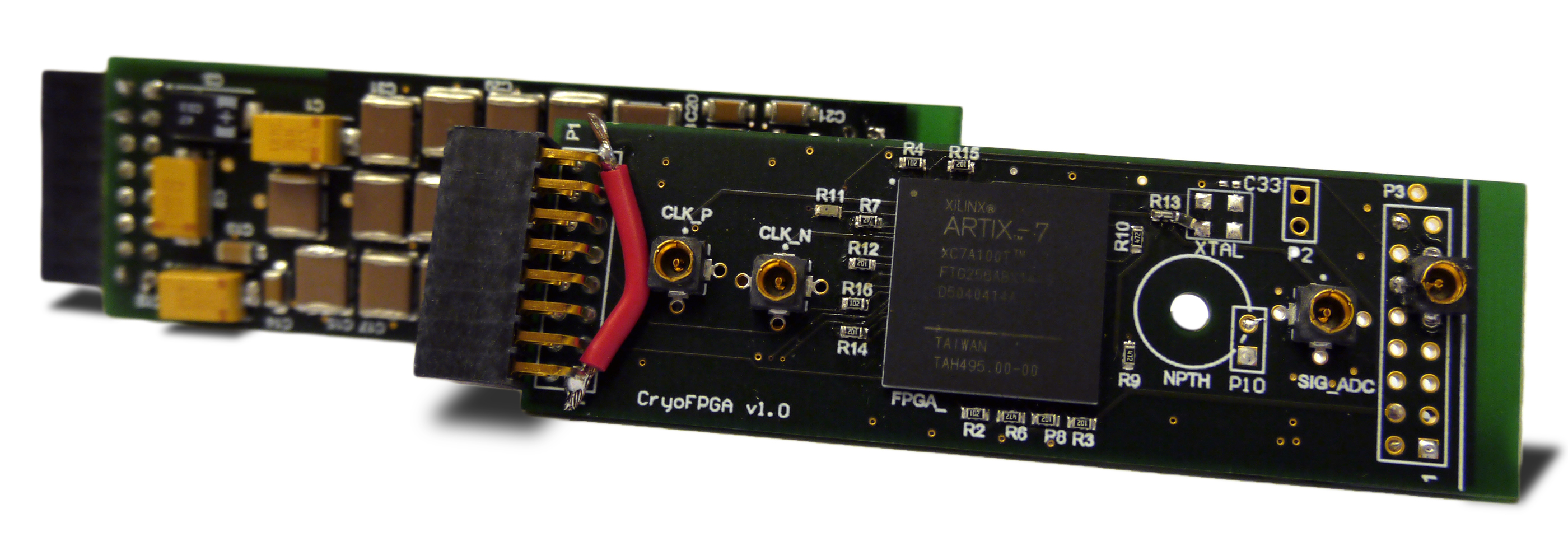}
			\put(0,0){\includegraphics[width=0.49\textwidth]{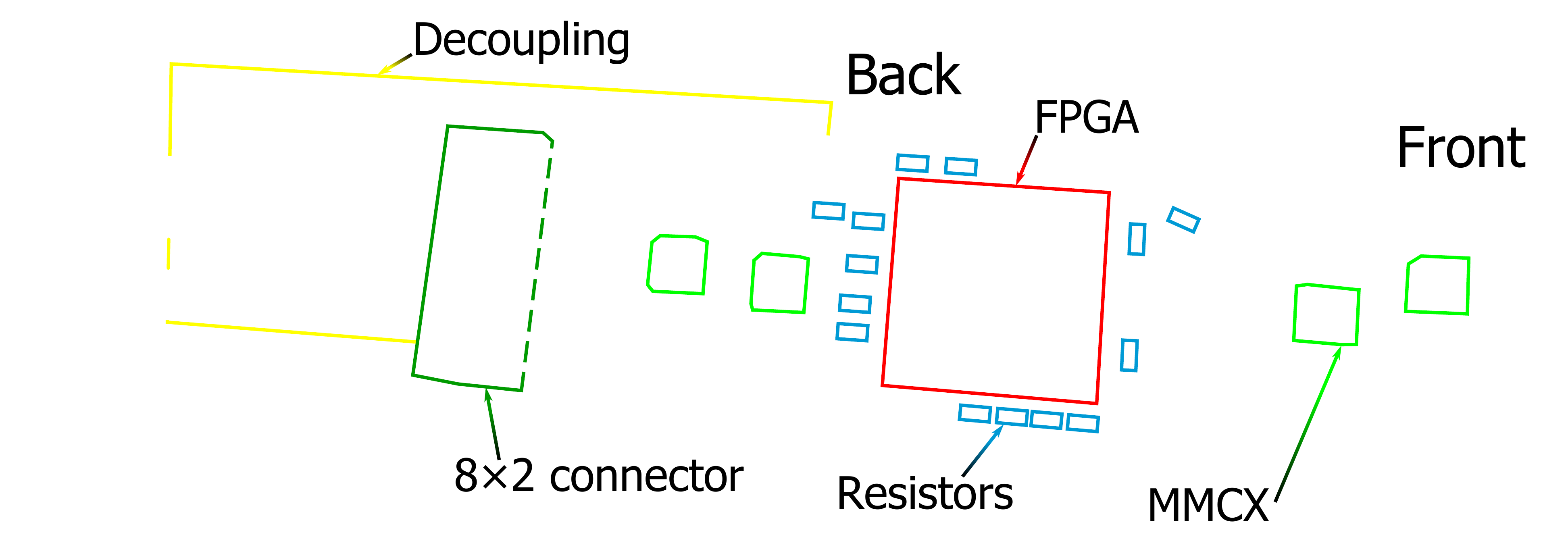}}
		\end{overpic}
		\vspace{-13pt}
		\caption{Cryogenic FPGA PCB. More details in appendix~\ref{sec:design_test}.}
		\label{fig:fpga_pcb}
	\end{figure}

\noindent A small PCB with a Xilinx Artix~7 (XC7A100T-2FTG256I) FPGA as the sole active component was designed for cryogenic testing. The PCB is shown in \autoref{fig:fpga_pcb}. Besides the FPGA, various passive components, most noticeably the decoupling capacitors, are present on board. More details on the PCB are presented in appendix~\ref{sec:design_test}.

Modern FPGAs, such as the Artix~7 from Xilinx, integrate a variety of different components, from simple look-up tables to advanced clock managers \cite{FPGA_blocks}. In \autoref{tab:fpga_working} an overview of the main FPGA components is given together with the performance change when operating at 4K. All those components are functional and their performance is comparable to room temperature. 

\begin{table}[b]
	\centering
	\scriptsize
	\caption{Overview of the working elements inside the CryoFPGA operating around 4K. Component details in \cite{FPGA_blocks}. The test procedure is detailed in \autoref{tab:fpga_working_extended}.}
	\begin{tabular}{m{1.6cm}|cm{5cm}}
		\hline
		{Module} & {Functional} & {Performance w.r.t. RT} \\
		\hline
		IOs        & \cmark \\
		LVDS 	   & \cmark \\
		LUTs       & \cmark & Propagation delay changes $<$ 5\%, jitter increases $<$ 15\% (11.8 to 13.5~ps) \\
		CARRY4     & \cmark & Propagation delay changes $<$ 2\%, jitter increases $<$ 3\% (11.5 to 11.8~ps) \\
		BRAM       & \cmark & No corruption in 10 test sets of 80~kB\\
		MMCM       & \cmark & Jitter reduction of roughly 20\% (11.4 to 8.9~ps) \\
		PLL        & \cmark & Jitter reduction of roughly 20\% (12.2 to 9.7~ps) \\
		IDELAYE2   & \cmark & Delay change of up to 30\%, jitter increase up to 50\% \\
		Temperature diode   & \cmark & Operating range 4K $-$ 300K \\
		\hline
	\end{tabular}
	\label{tab:fpga_working}
\end{table}

The delay change of look-up tables and carrychains is less than 5\%. While this is negligible in most applications, the change in logic speed is significant for the performance of both ADC and TDC.

\begin{figure}[t]
		\centering
		\includegraphics[width=0.4\textwidth]{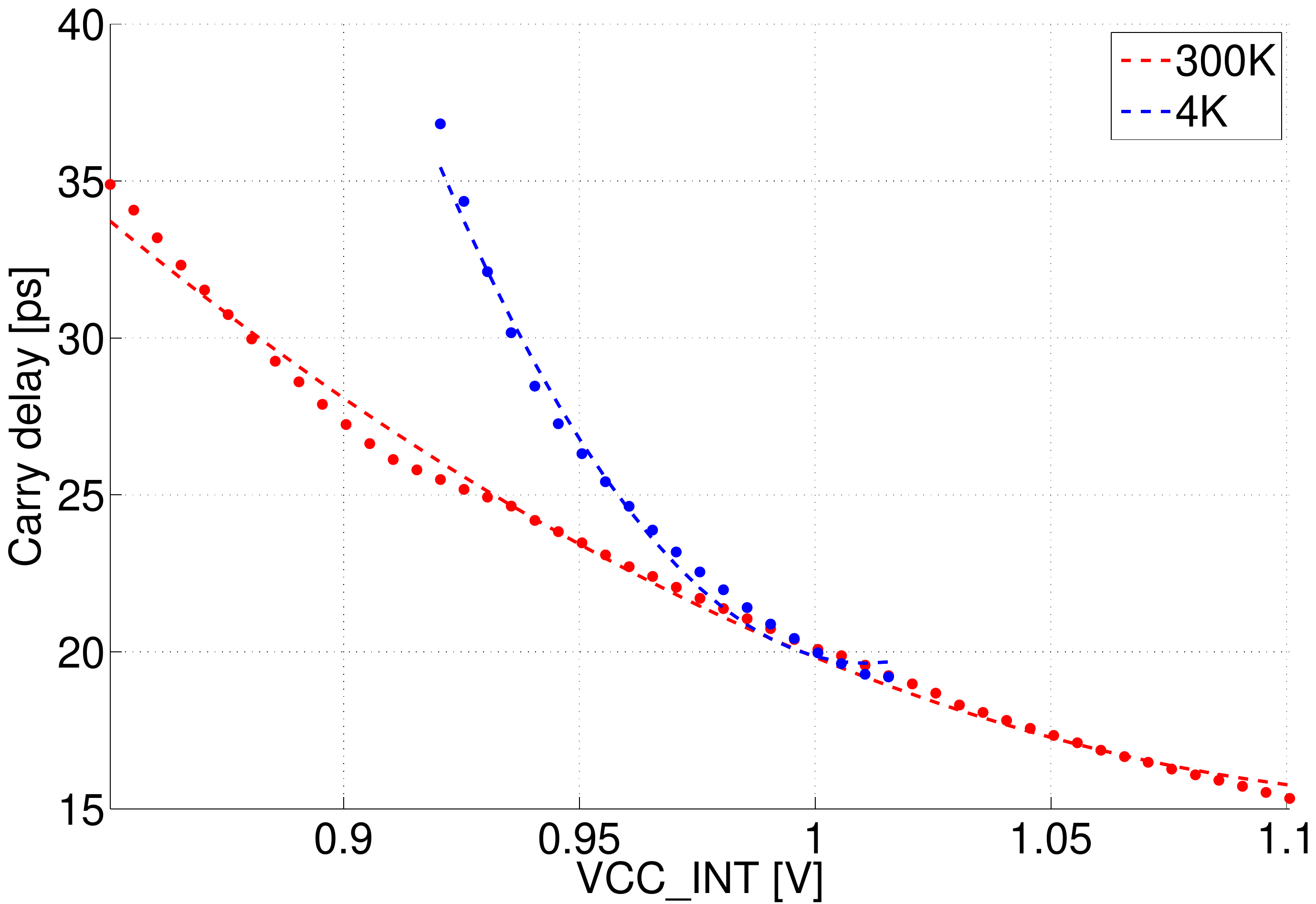}
		\caption{Delay of the carry elements in the carrychain versus FPGA voltage (VCC\_INT) over multiple temperatures.  }
		\label{fig:results_carries_resolution}
	\end{figure}

Therefore the main structure of those two circuits, i.e. the carrychain (used as delayline) is characterized more extensively over both FPGA internal voltage and temperature. \autoref{fig:results_carries_resolution} shows the average delay per carryblock versus the FPGA voltage for 300K and 4K. The delay change is significantly larger over the voltage range then over temperature, again signifying the performance stability over temperature. However, it also shows that the voltage must be very stable in order to achieve TDCs and ADCs that can be properly calibrated.

\begin{figure}[b]
		\centering
		\subfloat[]{\label{fig:oscillator_sweep}
			\includegraphics[width=0.2\textwidth]{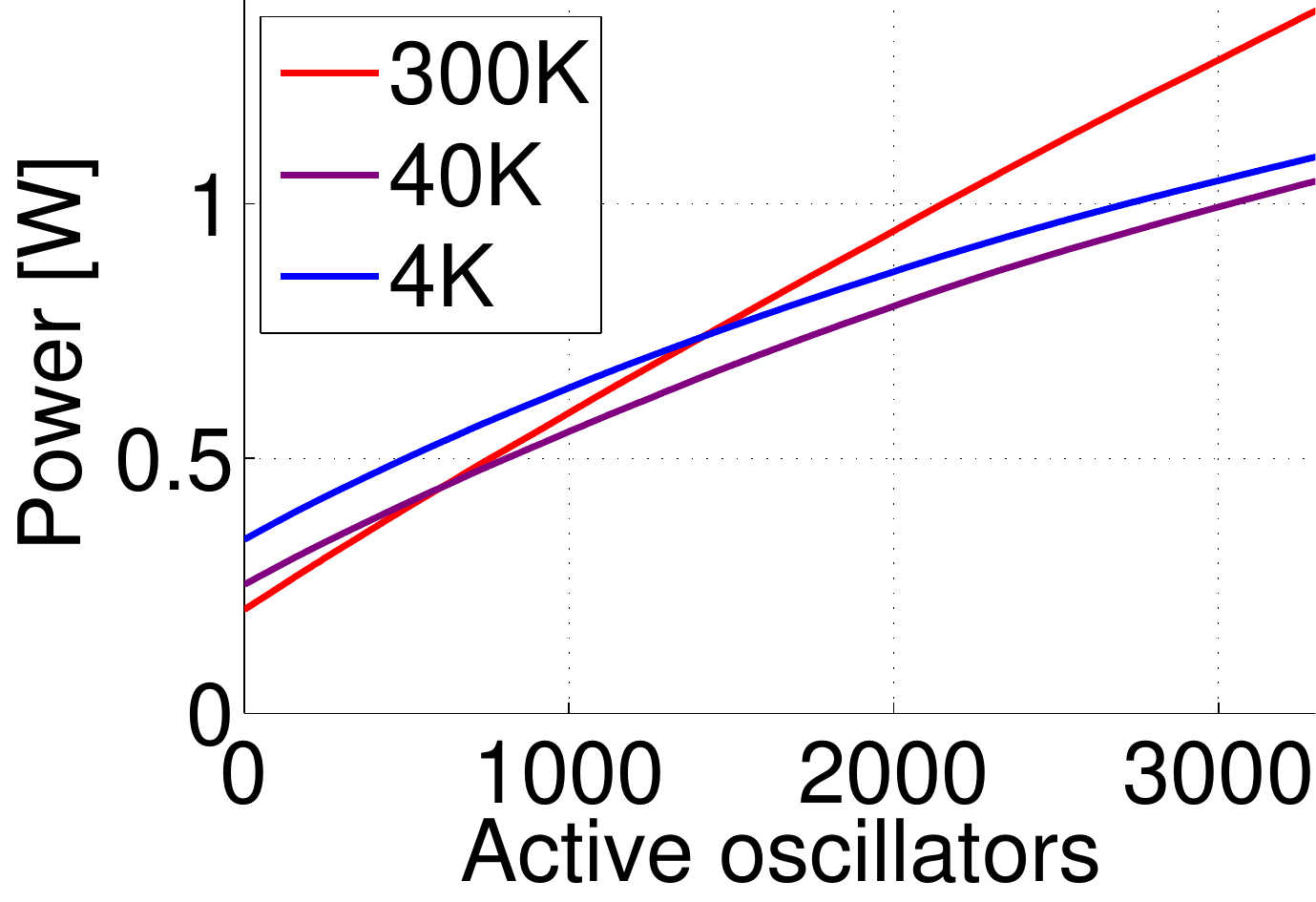}
		}
		\hfill
		\subfloat[]{\label{fig:current_sweep}
			\includegraphics[width=0.2\textwidth]{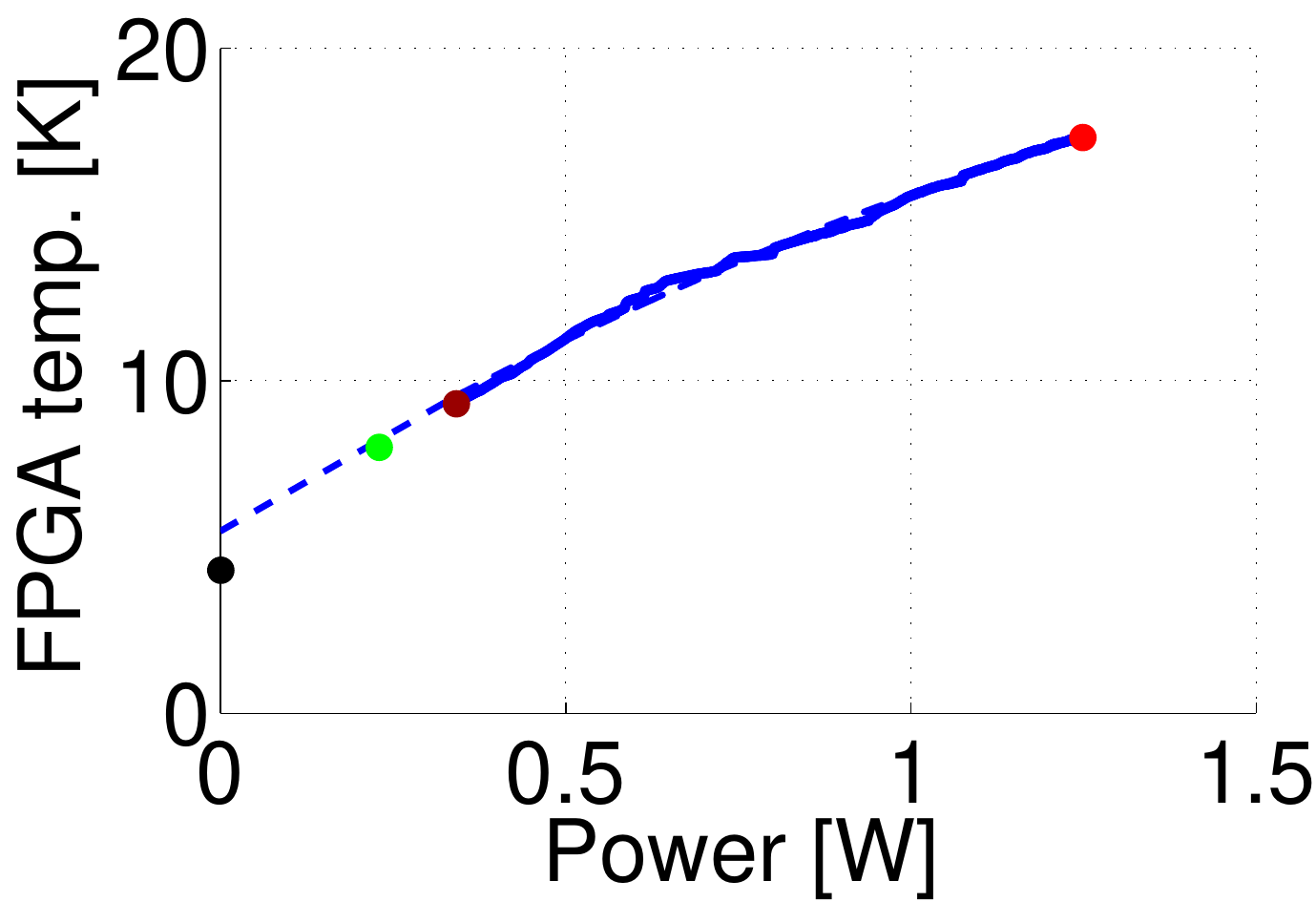}
		}
		\caption{\protect\subref{fig:oscillator_sweep} Power consumption of the FPGA while increasing the number of running oscillators in the FPGA fabric. \protect\subref{fig:current_sweep} FPGA power consumption versus the resulting die temperature. The coloured dots indicate (left to right): powered off FPGA, idle FPGA, 0 active oscillators, 4500 active oscillators. }
		\label{fig:oscillator_current_sweep}
	\end{figure}

To study self-heating effects and change in power consumption over temperature, a circuit was designed to be able to sweep the power consumption of the FPGA. The sweep was repeated at different temperature levels as shown in \autoref{fig:oscillator_current_sweep}\subref{fig:oscillator_sweep}.
A significant difference is observed in the response at the different temperatures. First, the idle power consumption increases from 83~mW at 300K to 228~mW at 4K. Since leakage is expected to decrease at lower temperatures, the increase of idle power is attributed to malfunctions of the support bias circuitry.

Secondly, a change in logic efficiency occurs. Namely, the average power per transition decreases and the power consumption starts to scale non-linearly, while lowering temperature. 
The average energy used per LUT decreases from 32 to 21~$\mu$W, i.e. the energy consumed per transition in a LUT decreases from 23 to 15~pJ. 
This implies a decrease in dynamic power consumption of over 30\%.

As shown in \autoref{fig:oscillator_current_sweep}\subref{fig:current_sweep}, the FPGAs junction temperature increases with power consumption from 4.3K (powered off FPGA) to 17.3K (at 1.25~W power consumption).
While the die temperature is elevated due to self-heating, the environment remained stable at a temperature of 4.2K ($\pm$0.1K). 

From the slope we can estimate a thermal resistance of the FPGA package of 8.5K/W, which is lower than the thermal resistance at room temperature in air (31K/W). \\

\begin{figure}[h]
		\centering
		\includegraphics[width=0.4\textwidth]{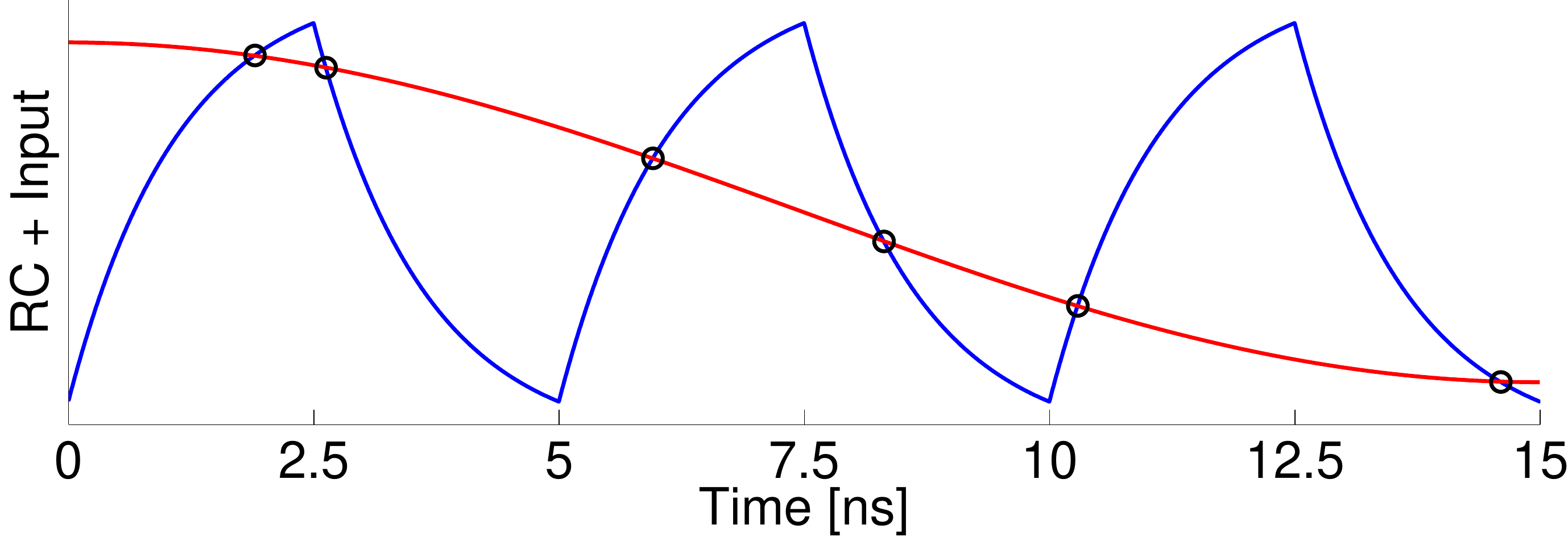}
		\caption{Basic principle of analog-to-digital conversion in the FPGA using an $RC$-filter to create a ramp from the clock output signal. The crossings of the input and the ramp are timestamped in a TDC.    }
		\label{fig:adc_principle}
	\end{figure}

\noindent One of the most important elements in the quantum error-correcting control loop is the analog-to-digital converter. Various attempts have been undertaken to optimize ASIC ADCs for cryogenic conditions, but reaching high performance is still challenging, as discussed before. Therefore, we opted to implement the ADC as a completely digital building block inside the FPGA. The operating principle is shown in \autoref{fig:adc_principle}\cite{Homulle2015}. A high speed clock signal is output on an FPGA pin connected, through a resistor, to an $LVDS$ capable input pin. Thanks to the resistor and the parasitic capacitance of the $LVDS$ buffer, an $RC$-ramp is created. The conversion is based on measuring the time required for the $RC$-ramp to cross the input signal, as done in a single-slope converter. The $LVDS$, used as comparator, switches as soon as the ramp voltage exceeds the input voltage and the $LVDS$ switching time is measured using a TDC. The performance of the TDC at cryogenic temperatures is given in appendix~\ref{sec:cryo_results}. 

With the use of multi clock-phase interleaving and calibration features, sampling rates, up to 1.2~GSa/s, above the clock frequency (400~MHz) are achieved. Furthermore, thanks to calibration, the ADC performance was more stable over temperature, process and voltage variations \cite{Visser2015}. \\

\noindent The performance of a 1.2~GSa/s ADC, created from interleaving 6 different phases of a 200~MHz clock, is summarized in \autoref{tab:adc_performance}. The FPGA is operating at a slightly elevated junction temperature of 15K, while the decoupling could not sustain the high power demand (of almost 0.9~W) at 4K to reach optimal performance. As the ADC is calibrated at both operating temperatures, the non linearities can be seen to be only slightly larger at cryogenic temperatures.

\begin{figure}[h]
		\centering
		\includegraphics[width=0.4\textwidth]{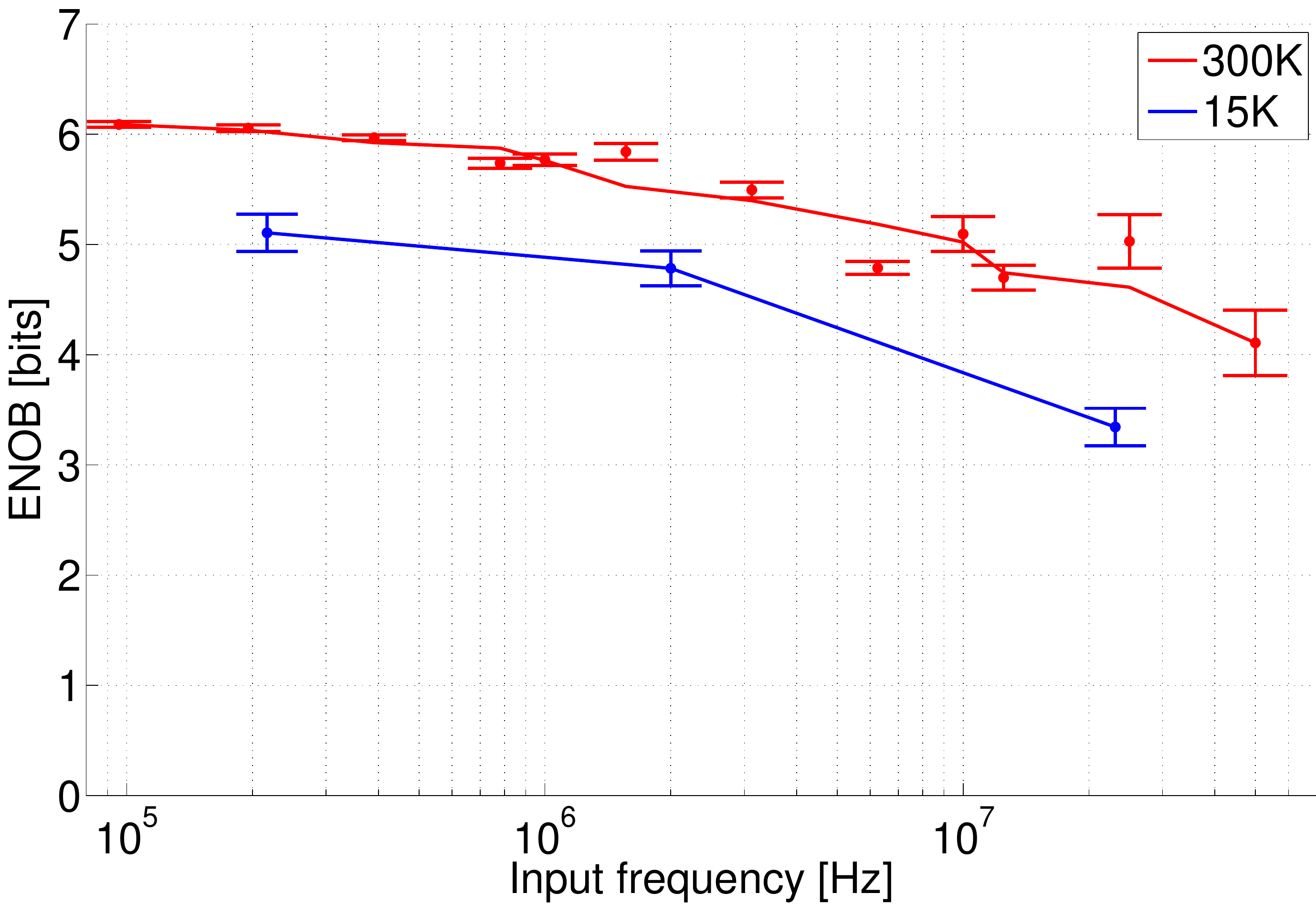}
		\caption{Effective number of bits versus input signal frequency at room temperature and 15K for the FPGA ADC sampling with 1.2~GSa/s. The length of the error bars is 2$\sigma$, the variation over 10 measurements. The solid lines are merely a guide for the eye. }
		\label{fig:adc_enob}
	\end{figure}

In \autoref{fig:adc_enob} the performance of the ADC is plotted in terms of ENOB over different signal input frequencies for both 300K and 15K. The ENOB is lowered with roughly one bit at cryogenic temperatures due to decreased decoupling stability and increased IDELAYE2 jitter. Although one bit has to be sacrificed while going down towards 15K, the performance is still in line with the requirements for the qubit control loop, which demands a high sampling rate rather than a high voltage resolution.

	\begin{table}[bt]
	  \centering
	  \footnotesize
	  \caption{Summary of calibrated ADC performance at 300K and 15K. The results are achieved after merging the 6 phase interleaved channels. }
		\begin{tabular}{lc|cc}
		\hline
		Temperature & [K] & 300   & 15 \\
		\hline
		Sampling rate & [MHz] & 1.2~GSa/s   & 1.2~GSa/s \\
		Input range & [V] & [0.9-1.6]  & [0.9-1.6] \\
		ENOB & [bits] & 6.0  & 5.0 \\
		DNL & [LSB] & [-0.75 1.04] & [-0.85 1.04] \\
		INL & [LSB] & [-0.36 0.52] & [-0.68 0.77] \\
		\hline
		\end{tabular}
	  \label{tab:adc_performance}
	\end{table}

\section{Conclusions}
\noindent To conclude, a cryogenic all-digital FPGA based control platform for error-correcting loops has been implemented. The FPGA operates over a large temperature range from 4K to room temperature, without significant change in performance. Logic speeds showed very high stability by changing less than 5\% over the whole temperature range. All major components such as MMCM, PLL, BRAM, IDELAY2 are adequately stable over the complete range.

Digital logic, analog-to-digital converters, temperature sensors and oscillators have been shown to operate correctly also at base temperature, while being integrated in the FPGA. Such set of control electronics, to be completed as a future perspective with digital-to-analog converters, paves the way towards practical control at cryogenic temperature of fault-tolerant solid state quantum information processing.

\section*{References}
\bibliographystyle{aipnum4-1}
\bibliography{Arxiv2016_bibliography}

\newpage
\appendix

\section{Cryogenic FPGA design \& test}\label{sec:design_test}
\noindent The cryogenic test set-up was designed to be both functional and simple. We thus excluded all those components that might fail (and negatively influence the FPGA) and made the FPGA the sole active component under test.

\subsection{PCB design}\label{sec:pcb_design}
\noindent The design of the cryogenic FPGA PCB focusses on simplicity while allowing the possibility of making all relevant tests with the PCB. The PCB was limited in size to fit in the test set-up (see \autoref{sec:set_up}) and is shown in \autoref{fig:fpga_pcb}. 

\noindent The PCB was fabricated with standard FR4 dielectric and 4 layers (one ground plane and three signal / power planes). The FPGA (Xilinx Artix~7 series), the high-frequency connectors and some resistors  were placed on the front side. The back-side hosts only decoupling capacitors. A detailed list of the components on the PCB is presented in \autoref{tab:fpga_bom}.

\begin{table}[b]
  \centering
  \scriptsize
  \caption{Bill of Materials (simplified) of the FPGA PCB.}
    \begin{tabular}{l|ll}
	\hline
    Component  & Manufacturer & Part \\
	\hline
    FPGA       & Xilinx     & XC7A100T-2FTG256I \\
    Capacitors & Kemet      & C-series (Ceramic (NP0/COG, X8L)) \\
    Capacitors & Kemet      & T-series (Tantalum) \\
    Capacitors & Panasonic  & TQC-series (Polymer) \\
    Resistors  & Panasonic  & ERA-sries (Metal film) \\
    Connectors & Amphenol   & MMCX \\
    Connectors & Samtec     & 8$\times$2 female connector \\
	\hline
    \end{tabular}
  \label{tab:fpga_bom}
\end{table}

\noindent In particular, we chose the FPGA XC7A100T-2FTG256I due to its small size, low power consumption (28~nm process node) and industrial grade (higher standard temperature range down to -40$^\circ$C). It features over $100k$ logic cells (in $16k$ slices), up to 170 user IOs and measures 17$\times$17~mm. Capacitors were chosen after selecting the best materials for cryogenic conditions and their availability in various capacitor values. Low permittivity materials, such as ceramic NP0/COG materials, are generally well behaved at 4K \cite{Teyssandier2010,Kirschman2014}. However, since they are only available in 0.47~$\mu$F format, tantalum polymer was used for the bigger capacitors. As the FPGA is powered over long cables (over 4~meters), special care was taken in minimizing the total capacitors effective series resistance and inductance (ESR and ESL). ESR effects the capacitors ability to quickly source and sink current, especially in fast current switching applications.
According to \cite{Teyssandier2010} not only the capacitance can drop significantly at cryogenic temperatures, but also the ESR can increase up to 1.000 times, stressing the importance of selecting the appropriate capacitor types for operation in a cryogenic environment. 
For high frequency signals MMCX connectors were selected for their small size. 

A detailed summary of the used connections to the FPGA board is given in \autoref{tab:fpga_connections}. For communication both JTAG (programming) and UART were used, occupying most of the user IO pins available on the board.
For the power supplies, additional cables were added to avoid a large $IR$ drop in the shielded cables. Special care has been taken to ensure a proper ground of the complete set-up, by shorting the cable shield terminals both at the instrument side (room temperature) and at the PCB side (deep cryogenic temperature). \\

\begin{table}[tbp]
  \centering
  \scriptsize
  \caption{Connections towards the FPGA PCB. }
    \begin{tabular}{lcl}
	\hline
    Connector  & Occupied pins & Purpose \\
	\hline
    Shields    & 20$^*$         & Ground \\
    DC         & 3          & Voltages (VCC\_INT, VCC\_O and VCC\_AUX) \\
    DC         & 4          & External temperature sensor (DT-670) \\
    8$\times$2 & 4          & JTAG (TDI, TDO, TMS, TCK) \\
    8$\times$2 & 3          & UART (RX, TX, CTS) \\
    8$\times$2 & 5          & Voltages + ground \\
    8$\times$2 & 2          & Internal temperature diode \\
    8$\times$2 & 2          & Debug \\
    MMCX       & 2          & Clock (differential) \\
    MMCX       & 1          & ADC input signal \\
    MMCX       & 1          & High frequency debug \\
	\hline
    \end{tabular}
	\vspace{2pt}
	\newline $^*$ 16+4 coaxial shields
	\vspace{2pt}
  \label{tab:fpga_connections}
\end{table}

\begin{figure}[tbp]
		\centering
		\includegraphics[width=0.44\textwidth]{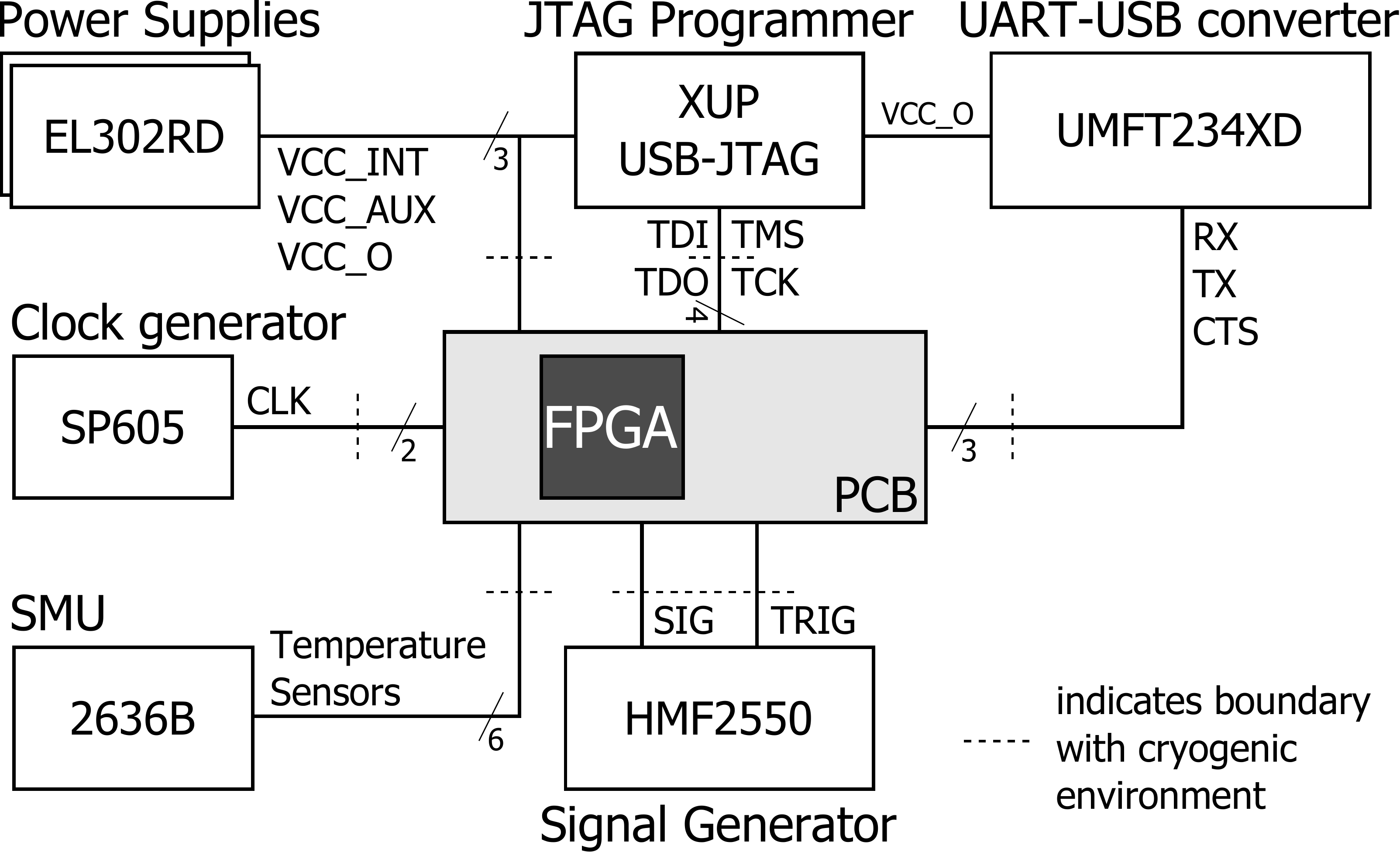}
		\caption{Overview of the equipment used outside the cryogenic environment to control and operate the FPGA. The ground is not indicated, but is common for all equipment and the PCB.  }
		\label{fig:cryo_boundary}
	\end{figure}

\noindent The measurement set-up is depicted in \autoref{fig:cryo_boundary}. For the power, two power supplies from Aim-TTI Instruments (EL302RD) were used, generating 1, 1.8 and 2.5~V. Programming and communication with the FPGA was done with a Digilent JTAG controller (XUP USB-JTAG) and an UART-USB dongle (FTDI UMFT234XD), respectively. The clock signal was provided differentially from an SP605 (Xilinx Spartan~6 evaluation kit) and was set to 100~MHz. Additionally a Rohde \& Schwarz HMF2550 signal generator has been used to provide the ADC test input. For accurate temperature measurements, a Keithley source meter unit (2636B), was added to the system. It provides a precision current bias for the FPGAs internal temperature-measuring diode and for an external reference sensor (Lake Shore Cryotronics, DT-670 silicon diode) mounted in thermal contact to the FPGA. Both diodes were biased with a current of 10~$\mu$A and the diode voltage is read as a function of temperature. 

\subsection{Test set-up}\label{sec:set_up}
\begin{figure}[tbp]
		\centering
		\includegraphics[width=0.17\textwidth]{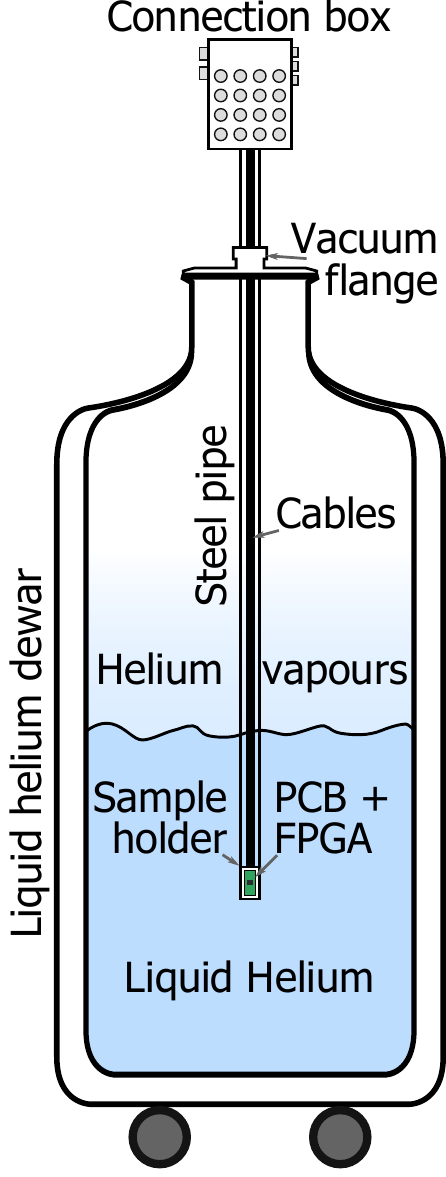}
		\caption{Schematic depiction of our cryogenic test set-up. The sample is mounted on a long steel pipe which is shifted into a helium dewar to reach the liquid helium. The connections to the sample are made with long (standard) cables that are attached to a break-out box on the top. There are 16 shielded connections to a 8$\times$2 pin header, 4 shielded connections to high frequency MMCX connectors and up to 9 additional DC wires for a temperature sensor and the various supply voltages. }
		\label{fig:cryo_setup}
	\end{figure}

\noindent In order to test the FPGA in deep cryogenic conditions, a set-up was built which uses a dip-stick to immerse the PCB in liquid helium. A schematic cross-section of the set-up is shown in \autoref{fig:cryo_setup}. The dip-stick consists of a steel pipe (2~m), a break-out box for cables placed on the top end and a sample holder at the low end of the pipe. The pipe limits the size of the PCB (diameter of 1.5~inch) and the number of cables. In total, 16 shielded cables (standard RG174/U coaxial wires with a 0.5~mm diameter core) are used for low frequency connections, 4 high frequency cables are connected to the MMCX and up to 9 additional cables for the temperature sensor and power supplies. \\

\noindent As the pipe is able to move up and down through a vacuum flange, the temperature can be shifted from 4K in the liquid phase to around 200K in helium vapours at the top of the dewar. As the sample is immersed in either the gas or the liquid helium, the cooling process is relatively fast and the cooling power relatively high compared to a vacuum refrigerator.

\section{Cryogenic FPGA results}\label{sec:cryo_results}

\begin{table*}[t]
	\centering
	\footnotesize
	\caption{Overview of the working elements inside the CryoFPGA operating around 4K.}
	\begin{tabular}{l|cm{6cm}m{6cm}}
		\hline
		{Module} & {Functional} & {Test} & {Performance w.r.t. RT} \\
		\hline
		IOs        & \cmark \\
		LVDS 	   & \cmark \\
		LUTs       & \cmark & LUTs connected to form oscillator of approximately 100~MHz at RT & Oscillation frequency changes $<$ 5\%, jitter increases $<$ 15\% (11.8 to 13.5~ps) \\
		CARRY4     & \cmark & Carrychains connected to form oscillator of approximately 100~MHz at RT & Oscillation frequency changes $<$ 2\%, jitter increases $<$ 3\% (11.5 to 11.8~ps) \\
		BRAM       & \cmark & Transfers of 8~kB (write \& read) & No corruption in 10 test sets of 80~kB\\
		MMCM       & \cmark & 100~MHz differential input clock multiplied by 10 and divided by 20 to 50~MHz single ended output & Jitter reduction of roughly 20\% (11.4 to 8.9~ps) \\
		PLL        & \cmark & 100~MHz differential input clock multiplied by 10 and divided by 20 to 50~MHz single ended output & Jitter reduction of roughly 20\% (12.2 to 9.7~ps) \\
		IDELAYE2   & \cmark & IDELAYE2 elements connected to form a tunable oscillator (output frequency variable 13-70~MHz) & Delay change of up to 30\%, jitter increase up to 50\% \\
		\hline
	\end{tabular}
	\label{tab:fpga_working_extended}
\end{table*}

\subsection{Extended evaluation of FPGA functionality \& performance}\label{sec:basic_testing}
\noindent A test procedure was defined to sequentially test the most vital FPGA components. As each component requires a different test procedure, different VHDL programs were designed and programmed on the FPGA through the JTAG interface. It should be noted that programming was working over the complete temperature range, allowing us to test the different components separately in the cryogenic environment. However the communication speed had to be set below 1~MHz, for both JTAG and UART, due to not proper termination of the coaxial lines causing too strong reflections at higher frequencies. 

In \autoref{tab:fpga_working_extended} an overview of the main FPGA components is given together with the test procedure and the performance change when cooling to 4K.

\begin{figure}[b]
		\centering
		\includegraphics[width=0.4\textwidth]{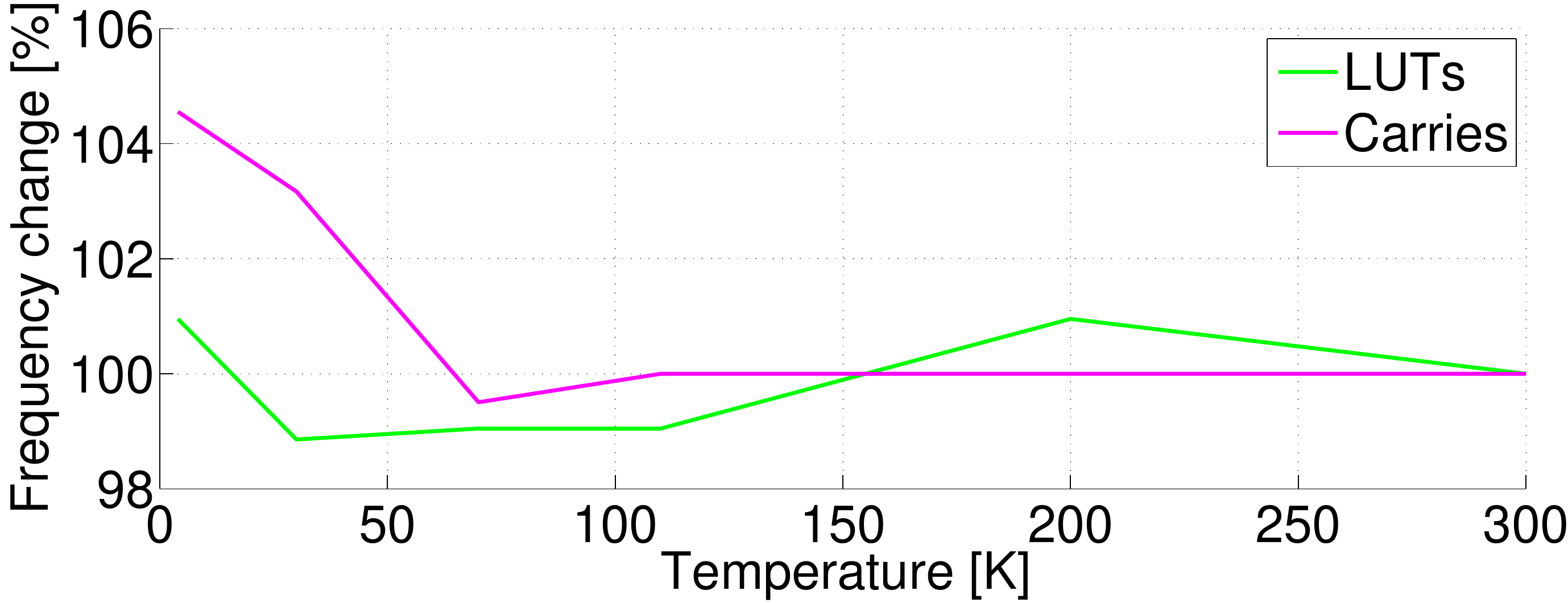}
		\caption{Change in oscillation frequency of oscillators build from look-up tables and carrychains respectively over a temperature sweep from 4K up to 300K. }
		\label{fig:results_lutscarries_freq}
	\end{figure}

\autoref{fig:results_lutscarries_freq} shows that the change in oscillation frequency of two different oscillators (both tuned to be around 100~MHz) implemented with either look-up tables and carrychains is below 5\%.

The operating voltage range of the FPGA changes significantly over temperature. The range is wider at room temperature (0.85~V - 1.1~V) than the specified range (0.95~V - 1.05~V). At low temperature, it reduces significantly on both ends (0.92~V - 1.02~V). \\

\noindent The FPGAs internal temperature diode was characterized against a precise external reference (Lake Shore Cryotronics, DT-670 silicon diode), which is calibrated down to 1.4K. During the test, the FPGA was switched off completely, to minimize self-heating effects.

	\begin{figure}[b]
		\centering
		\includegraphics[width=0.4\textwidth]{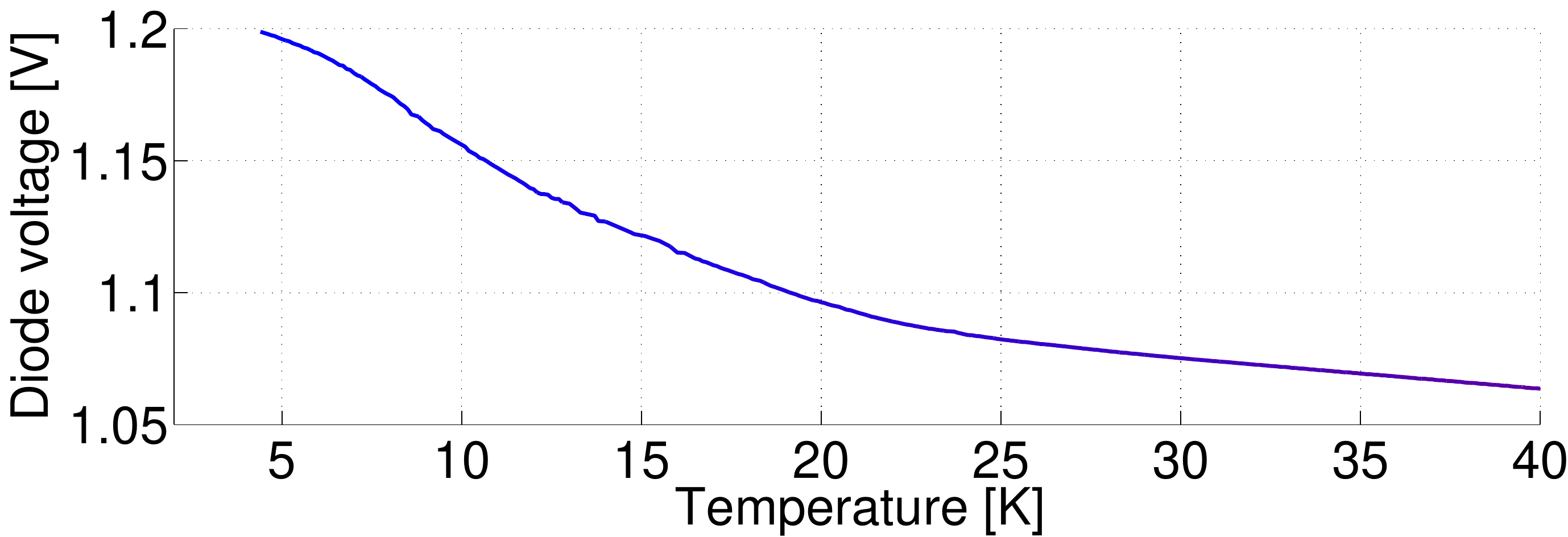}
		\caption{FPGA diode calibration against external reference diode, both diodes biased with 10~$\mu$A. Over 3.000 measurement points were used in order to recreate the diode curve. As some hysteresis was present in the measurement data, the curve is averaged out between temperature decrease and increase.  }
		\label{fig:diode_calibration}
	\end{figure}

The FPGA diode was calibrated with a 10~$\mu$A current and the voltage is stored as a function of the reference temperature (acquired by the reference sensor) as shown in \autoref{fig:diode_calibration}.

Idle power consumption for the various supplies at different temperatures is given in \autoref{tab:idle_currents}. 

	\begin{table}[tb]
	\centering
	\footnotesize
	\caption{Idle FPGA power consumption at different temperatures. Bias voltages for VCC\_INT, VCC\_AUX and VCC\_O are 1, 1.8 and 2.5~V respectively. }
		\begin{tabular}{c|ccc|c}
		\hline
		Temperature & VCC\_INT  & VCC\_AUX  & VCC\_O  & Power consumption \\
		~[K] & ~[mA] & ~[mA] & ~[mA] & ~[mW] \\
		\hline
		300   & 24    & 26    & 5     & 83 \\
		40    & 19    & 69    & 4     & 153 \\
		4     & 20    & 110   & 4     & 228 \\
		\hline
		\end{tabular}
	\label{tab:idle_currents}
	\end{table}

\subsection{Time-to-digital converters}\label{sec:tdc}
\noindent As the FPGA is fully functional (even at 4K) and performance is comparable to that of room temperature, complex and high-speed circuits were implemented and tested. As we do not expect any (negative) change in performance for normal digital logic operations, some more demanding circuits were implemented. A TDC was implemented using the carrychain with 200 delay stages (50 Carry4 blocks) and running at 400~MHz. The sampling clock was generated by a mixed mode clock manager (MMCM) module on the FPGA together with a 100 and 200~MHz clock for the remainder of the logic. Following the TDC, a histogrammer was implemented to store the generated data in a compressed format (16~bits of data for each of the 200 bins). Each histogram bin is incremented when the corresponding time stamp is measured in the TDC. The histogram was consecutively read out through the UART module and transferred to the host. 
The basic structure of the TDC is described by \cite{Favi2009, Visser2015}. The histogrammer units are detailed in \cite{Homulle2014}.

This circuit employs all basic components: LUTs, carrychains, block RAM, clock managers and IO delays. It not only validates the conclusions from \autoref{sec:cryo_results_basic}, but shows all components can work together reliably at low temperatures with TDC performance comparable to room temperature.
	
	\begin{table}[tbp]
	  \centering
	  \footnotesize
	  \caption{Summary of (uncalibrated) TDC performance at 300K and 4K. The FPGA was operating at the base voltage, VCC\_INT of 1~V. }
		\begin{tabular}{lc|cc}
		\hline
		Temperature & [K] & 300   & 4 \\
		\hline
		Sampling rate & [MHz] & 400   & 400 \\
		Resolution & [ps] & 20.1  & 20 \\
		DNL & [LSB] & [-1 4.6] & [-1 2.8] \\
		INL & [LSB] & [-2.6 3.3] & [-3.9 2.3] \\
		Jitter (average) & [LSB] &  0.9     & 1.6 \\
		\hline
		\end{tabular}
	  \label{tab:tdc_performance}
	\end{table}

\begin{figure}[h]
		\centering
		\subfloat[]{\label{fig:cryo_tdc_a}
			\includegraphics[width=0.4\textwidth]{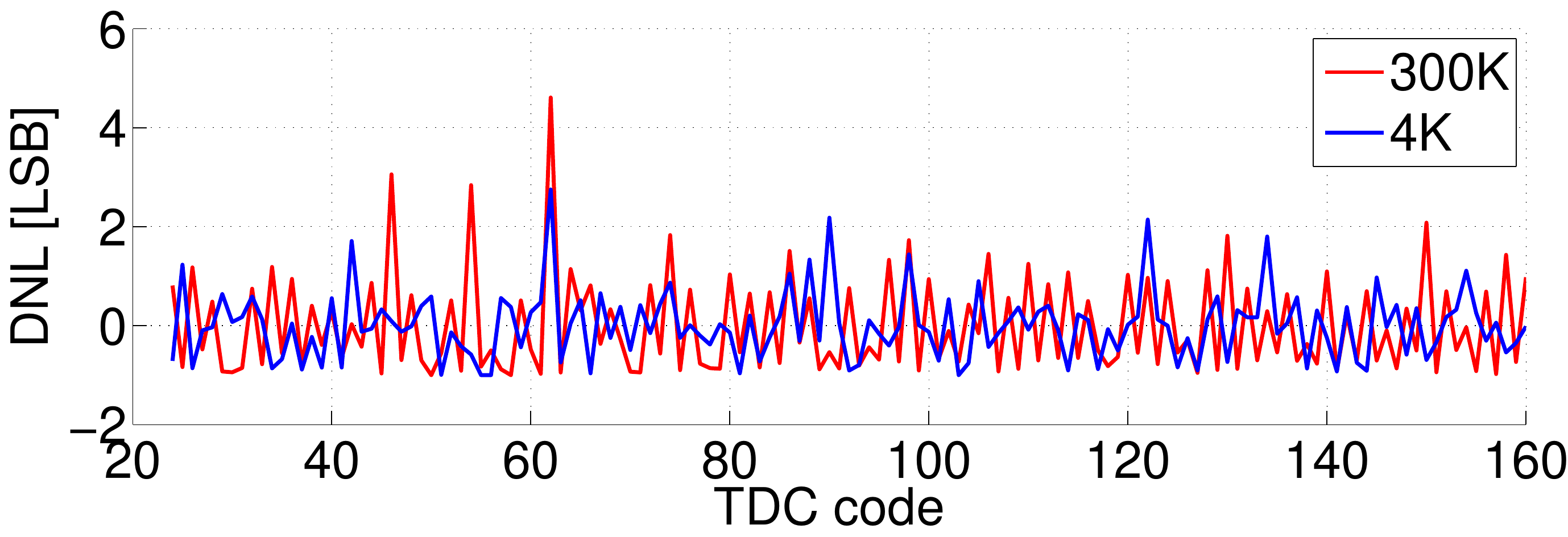}
		}\\
		\subfloat[]{\label{fig:cryo_tdc_b}
			\includegraphics[width=0.4\textwidth]{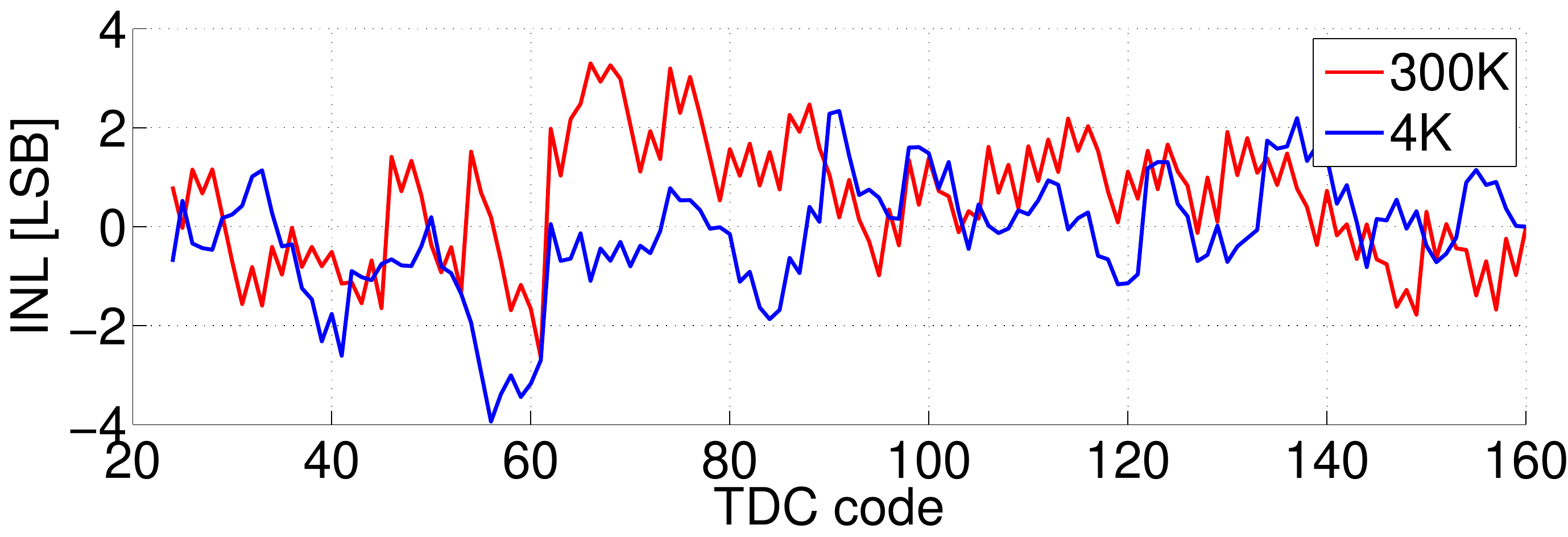}
		}\\
		\subfloat[]{\label{fig:cryo_tdc_c}
			\includegraphics[width=0.4\textwidth]{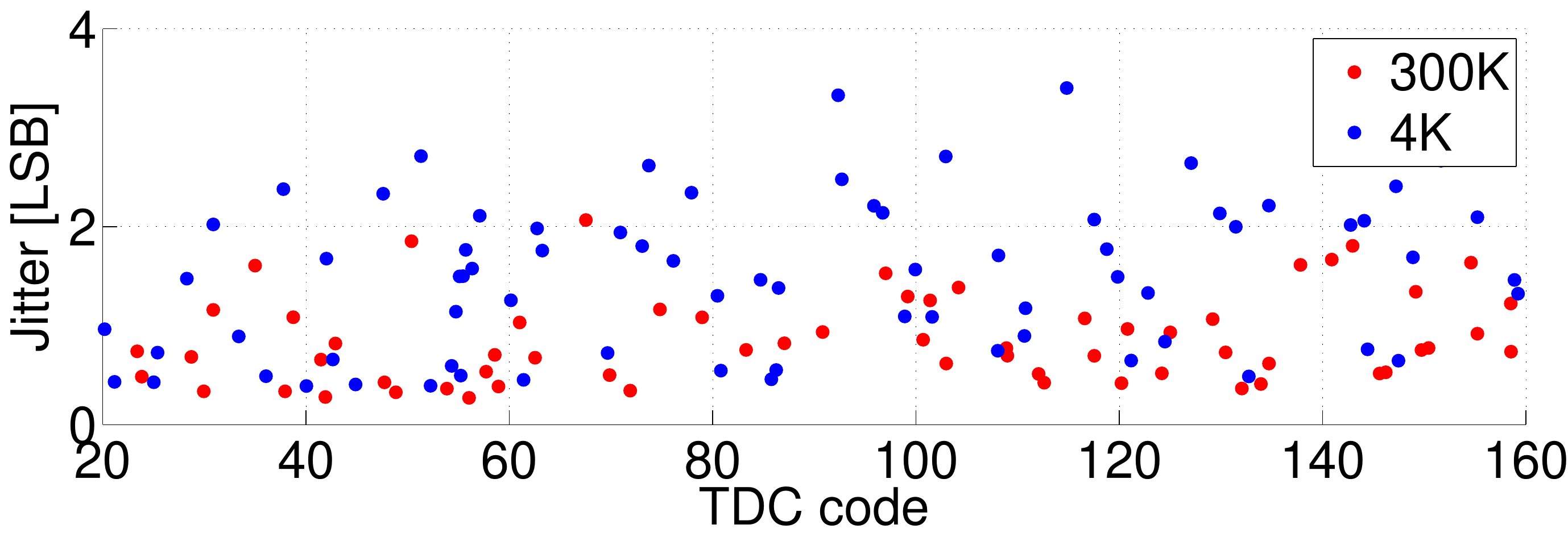}
		}
		\caption{
		Performance comparison of a time-to-digital converter operating in 300K and 4K. Both \protect\subref{fig:cryo_tdc_a} differential non linearity and \protect\subref{fig:cryo_tdc_b} integral non linearity are obtained from a density test with over 30 million counts in the histogram. \protect\subref{fig:cryo_tdc_c} jitter performance (1$\sigma$) of the TDC, captured with a synchronised clock shifted through the carrychain. All results obtained without calibration of the TDC. }
		\label{fig:cryo_tdc}
	\end{figure}

The most important parameters for a given TDC are its non linearity, jitter performance, resolution and sampling rate. A summary of the uncalibrated TDC performance at 300K and 4K is presented in \autoref{tab:tdc_performance}. The differential non linearity (DNL) and integral non linearity (INL) were obtained through a density test; over 30 million samples were used (from 400 integration runs). The extracted DNL and INL are shown in \autoref{fig:cryo_tdc}\subref{fig:cryo_tdc_a} and \subref{fig:cryo_tdc_b} respectively. Both room temperature and cryogenic non linearities are comparable. On contrary, the average jitter increases to 0.7~LSB (1$\sigma$)at cryogenic temperatures, thus w.r.t. room temperature. The jitter increase can be attributed mainly to the increase in jitter of the IDELAYE2 component as noted previously in \autoref{tab:fpga_working_extended}.

\end{document}